\documentclass[%
 reprint,
superscriptaddress,
 amsmath,amssymb,
 aps,
 pre,
]{revtex4-1}

\usepackage{url}

\usepackage[utf8]{inputenc} 
\usepackage{graphicx}
\usepackage{dcolumn}
\usepackage{bm}
\usepackage{xcolor}
\usepackage[bf]{subfigure}
\usepackage{amsmath}
\usepackage{tcolorbox}

\begin{document}

\title{On Hypercomplex Networks}

\author{\'Everton Fernandes da Cunha}
\affiliation{
S\~ao Carlos Institute of Physics, University of S\~ao Paulo, S\~ao Carlos, S\~ao Paulo, Brazil.
}

\author{Luciano da Fontoura Costa}
\affiliation{
S\~ao Carlos Institute of Physics, University of S\~ao Paulo, S\~ao Carlos, S\~ao Paulo, Brazil.
}

\begin{abstract}
The concept of `complexity' plays a central role in complex network science. Traditionally, this term has been taken to express heterogeneity of the node degrees of a therefore complex network. However, given that the degree distribution
is not enough to provide an invertible representation of a given network, additional complementary measurements of its topology are required in order to complement its characterization. In the present work, we aim at obtaining a new model of
complex networks, called hypercomplex networks -- HC, that are characterized by heterogeneity not only of the degree distribution, but also of a relatively complete set of complementary topological measurements including node degree, shortest path length, clustering coefficient, betweenness centrality, matching index, Laplacian eigenvalue and hierarchical node degree.  The proposed model starts with uniformly random networks, namely Erd\H{o}s-R\'enyi
structures, and then applies optimization so as to increase an index of the overall complexity of the networks. An optimization approach has been considered in terms of gradient descent. The complexity index expresses the dispersion of the several considered measurements. Several interesting results
are reported and discussed, including the fact that the HC networks define, as the optimization proceeds, a trajectory in the principal component space of 
the measurements that tends to depart from the considered theoretical models 
(Erd\H{o}s-R\'enyi, Barab\'asi-Albert, Waxman, Random Geometric Graph and Watts-Strogatz), heading to a previously empty space (low density of cases).  We observed that, after a considerably large number of optimization steps, peripheral branching tends to appear that further enhances the complexity of these networks.
\end{abstract}

\maketitle


\section{Introduction}

Network science has progressed a long way from its beginnings in about 2000, to the extent of becoming one of the most popular and widely applied areas in modern science (e.g.~\cite{costa2007characterization}). To a good extent, the importance of network science derives from two main facts: (i) being graphs, complex networks correspond to one of the most general discrete structures -- encompassing trees, lattices -- therefore allowing effective representation of any discrete system or problem; and (ii) the basic idea of complex network is accessible even to non-experts, allowing it to be understood and applied in a wide inter- and multidisciplinary manner.

Yet, despite the intrinsic simplicity of the idea of graph, the concept of \emph{complex networks} remains a challenge to be conceptually understood. The main issue here regards the term \emph{complex}. The most typically taken interpretation of this concept corresponds to a relative characterization with respect to some network models, especially Erd\H{o}s-R\'enyi (ER), acting as a \emph{simple} counterpart \cite{erdos1959random,erdos1960evolution} characterized by homogeneous connectivity. 

In particular, these simple reference models are so that the degree of their nodes can be relatively well-predicted from the average of all nodes degrees and controlled just by one parameter: the connection probability, \textit{p}. Thus, in a sense a simple graph is characterized by degree regularity or homogeneity, having a regular graph as its limit prototype. In this sense, uniformly random networks such as ER can therefore be understood as being statistically regular. 

Compared to these regular counterparts, network models such as Barab\'asi-Albert (BA) are understood as being \textit{complex}, because the degree of the nodes in this type of network is heterogeneous enough so that the properties of the network cannot be properly predicted from the overall average degree (e.g.~\cite{costa2007characterization,barabasi1999emergence}). As a matter of fact, the degree distribution of the BA model is scale free, implying the existence of nodes, the so-called hubs, with degree substantially larger than the average degree.

\begin{figure}[!t]
\centering
\includegraphics[width=6cm]{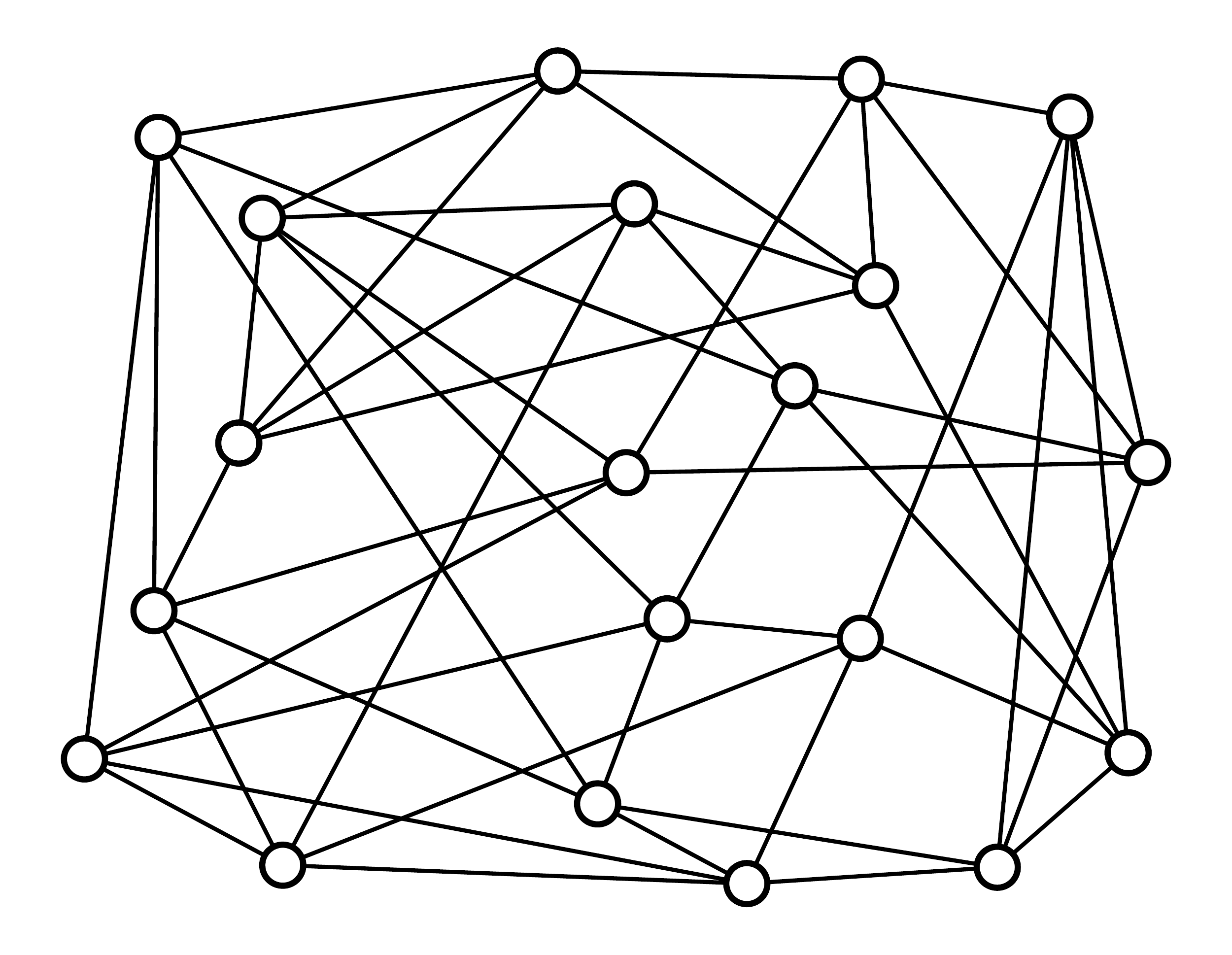}
\caption{Example of a network that, though perfectly regular in the sense that
each of its $N=20$ nodes have the same degree, $k=5$, cannot be intuitively
considered to have a simple overall connectivity structure.}
\label{fig:complexisregular}
\end{figure}

It has been shown (e.g.~\cite{kim2008complex,costa2018complex}) that degree heterogeneity is not enough to characterize a complex network, in the sense that it is possible to have full regular networks that nevertheless exhibit intricate topology. An example of such network is depicted in Figure~\ref{fig:complexisregular}.

Thus, the adjective \emph{complex} would in fact require heterogeneity of every possible topological measurement, or, at least, more than just degree distribution. 

This more comprehensive way to understand complex network has many important implications, of which we highlight the following two: (i) how to measure the complexity in a broader sense? (ii) what would be the most complex networks that can be obtained?

Our main goal in this work is to investigate these questions. For that, we propose a complexity index that corresponds to the average of the standard deviation of several measurements with a proper correction incorporating the measurements in a comparative manner. Then, we propose a algorithm to maximize the complexity index of a graph having flexible parameters allowing a range possibilities of networks with high complexity.

\section{Materials and methods}\label{sec:mat} 

Proposing a complexity coefficient based on the diversity of topological measurements creates a necessity to select a group of topological 
features that can provide a reasonably comprehensive characterization
of the network structure (e.g.~\cite{costa2007characterization}).  However, global proprieties have been avoided because they just give us information about the entire graph instead of its possibly heterogeneous parts. For this purpose, ten topological measurements were estimated for in the network, including: (i) node degree; (ii) average and (iii) standard deviation of the shortest path length for each par of nodes; (iv) clustering coefficient of each node; betweenness centrality for (v) nodes and (vi) edges; (vii) matching index for every pair of nodes that shared an edge; (viii) Laplacian eigenvalues; and (ix) second and (x) third hierarchical degree of each node. Considering the networks to be unweighted and non-directed, we can understand theses measurements as follows:

\textbf{Node degree:} the number of the neighbors of a node, quantifying how well a node is connected to other nodes in the graph;

\textbf{Shortest path length:} the smallest sequence of adjacent nodes between two nodes in a graph is the \textit{shortest path length} (or \textit{geodesic path length}). We consider the average and standard deviation of this measurement taken between each node and all other nodes in the network.

\textbf{Clustering coefficient:} information of how well the neighbors of a node are interconnected can be provided by the \textit{clustering coefficient} \cite{watts1998collective}. It can be defined as:

\begin{equation}
C_{i} = \frac{N_{\Delta}(i)}{N_3(i)};
\end{equation}

\noindent where $N_{\Delta}(i)$ is the number of all different pairs of nodes that are neighbors of node $i$ that are connected each other. $N_3(i)$ is the number of all possible pairs of node $i$, calculated as $N_3(i) = {k_i(k_i-1)}/{2}$. This measurement is taken for each of the network nodes;

\textbf{Betweenness centrality}: This measurement provides an indication of how many shortest paths go through the network nodes and edges \cite{freeman1977set}. It can be expressed as:

\begin{equation}
B_{u} = \sum_{ij} \frac{\sigma(i, u, j)}{\sigma(i, j)};  
\end{equation}
\noindent where $\sigma(i, u, j)$ presents how many shortest paths exist between the distinct nodes $i$ and $j$ that the node or edge, $u$, is included; $\sigma(i, j)$ is the total value of the geodesic paths between $i$ and $j$ and the sum is done over all pairs $i$, $j$ where $i \ne j$.
This measurement was taken for each node and each edge in the network;

\textbf{Matching index:} This measure indicates the similarity between two linked nodes (e.g.~\cite{hilgetag2002computational,arruda2018representation}) in the sense of the relative number of shared neighbors. It can be expressed as:

\begin{equation}
\mu_{ij} = \frac{ \sum_{k \ne i,j}a_{ik}a_{jk}}{ \sum_{k \ne j}a_{ik} + \sum_{k \ne i}a_{jk}};  
\end{equation}
\noindent where $a_{ij}$ is a element of a adjacency matrix and when $a_{ij}=1$ means that the nodes $i$ and $j$ are connected;

\textbf{Laplacian eigenvalue:} information about the topology of a network can be obtained from spectral methods based on the analysis of eigenvalues and eigenvectors from adjacency matrix of the graph \cite{seary1995partitioning,newman2006finding}. Specifically, we use the \textit{Laplacian} matrix, defined as: 

\begin{equation}
L = D - A;  
\end{equation}

\noindent where D is the diagonal matrix of node degrees and A is the adjacency matrix. Using that, it is obtained the \textit{Laplacian eigenvalues} of the network. As the graphs considered here is non-directed and unweighted, the values of this measurement are positives and just have real parts \cite{newman2018networks}.

\textbf{Hierarchical node degree:} Given a node in a network, it is possible to generalize the concept of its degree by considering not only its first neighborhood, but also other subsequent neighborhoods (e.g.~\cite{costa2006hierarchical, costa2007characterization}). In this work, we take the second and third hierarchical degrees of each node in the networks.

\subsection{The Complexity Index}
As argued in~\cite{costa2018complex}, the complexity of a given network does not limit itself to heterogeneity of node degree, and should also encompass the diversity of other topological measurements. In a sense, the more dispersed are all possible such measurements, the more heterogeneous and, therefore, complex the network can be understood to be. 

In order to quantify the overall complexity of a given network, it is important to have
some complexity index that takes into account several of the respective topological features
of the network.  In the present work, we propose one overall complexity index, henceforth
referred to as $CI$, that is derived from the standard deviations of several measurements, $\sigma_m$. Indeed, each of these standard deviations provides on itself some indication of
the complexity of the network regarding that respective feature. 
Observe that these measurements may be respective to each node or each edge.
Therefore, first we obtain several such standard deviations characterizing the heterogeneity of the topological properties of a given network, one respective to each of the adopted topological measurements. 

One important point when trying to quantify the complexity of a given network is that
this concept is relative to other networks. In other words, one type of network can be said
to be more complex than other types, while it remains a difficult problem to express its
complexity in an absolute manner.  Therefore, in the present work we take into account,
while estimating the complexity index of a given network, not only that network, but a
a whole set of reference networks. The choice of these reference networks is not fixed and
can be determined in terms of each specific research. For instance, in the present work
we will consider the complexity of the HC networks with respect to Erd\H{o}s-R\'enyi, Barab\'asi-Albert, Waxman, Random Geometric Graph and Watts-Strogatz.

Because each topological measurement can have a distinct scale of variation (for instance, the
average degree is typically larger than 1, the clustering coefficient varies from 0 to 1), it
becomes necessary to normalize the values of the considered measurements. Here, we resource to
the standardization of the each adopted measurement $m$, which can be obtained as follows:

\begin{equation}
z_m = \frac{m - \mu_m}{\sigma_m};
\end{equation}

\noindent where $\mu_m$ and $\sigma_m$ are, respectively, the average and standard deviation of measurement $m$.

Observe that the standardization is performed considering not only the network whose complexity
index is being calculated, but also all the other networks taken as reference.

The standard deviations of each measurement obtained for the network of interest
after standardization now need to be combined in order to obtain a single complexity index.
Ideally, a maximally complex network would have equally large standard deviations for all considered
measurements. Consequently, any difference between the standard deviations will reduce the
overall network complexity, as understood in the present work.

\begin{figure}[!t]
\centering
\includegraphics[width=6cm]{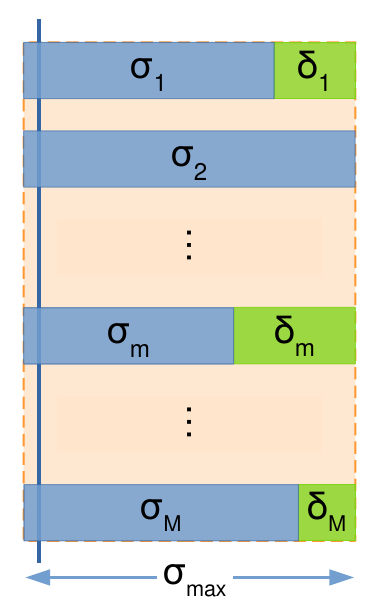}
\caption{Construct taken into account for defining the complexity index. Each line corresponds
to the standard deviation of each of the standardized measurements observed for the
network whose complexity is to be quantified. The maximum standard deviation 
defines a bounding rectangle from which the standard deviation differences 
$\delta_m$ can be obtained. The maximum relative complexity of 1 is understood to take
place when no such differences are observed.}
\label{fig:CI}
\end{figure}

Let's represent the largest standard deviation (please refer to Figure \ref{fig:CI}) by $\sigma_{max}$,
i.e.:

\begin{equation}
\sigma_{max} = \max\limits_{m=1,M}( \sigma_m ).
\end{equation}

Now, we define the quantity $\beta = M \, \sigma_{max}$ as corresponding to the `maximum
effective variance' possible for the $M$ measurements.

For each measurement $m$, we can define the difference between the maximum standard deviation
and the observed respective standard deviation as $\delta_m = \sigma_{max} - \sigma_m$.
Thus, the sum $\Delta$ of differences $\delta_m$ can be expressed as:

\begin{equation}
\Delta = \sum_{m=1}^{M}\delta_m.
\end{equation}

A preliminary index $\alpha$ can now be defined as:

\begin{equation}
\alpha = \frac{\beta - \Delta}{\beta}.  
\end{equation}

This index expresses the relative complexity of the network of interest. It will be maximum, $\alpha = 1$,
when $\Delta =0$, indicating that all measurements have the same dispersion. However, any
difference $\delta_m$ observed for each measurement $m$ will contribute to increase $\Delta$ value, and therefore, to reduce the
value of $\alpha$. In particular, the smallest relative complexity will be obtained when
all standard deviations, except one of them, will be zero. In this case, we have
$\Delta = \sigma_{max} = \beta$, implying $\alpha = 0$. Therefore, we have that
$0 \leq \alpha \leq 1$.

Now, we can proposed an absolute overall complexity index, $CI$, as being equal to the relative
complexity index times the average standard deviations observed for the network of interest:

\begin{equation}
CI = \alpha \left< \sigma \right>.  
\end{equation}

In order to better understand the difference between the relative and absolute indices,
let's consider two networks, all of which having all standard deviations identical to 10
and 100, respectively. Though both these networks will have maximum relative complexity 
$\alpha$, the second network is plainly more complex because of the larger values of its
measurement dispersions. This difference is properly reflected in the absolute complexity 
index, which will result much larger for the second situation.

\subsection{Seeking for complexity}

\begin{figure}[!t]
\centering
\includegraphics[width=7.5cm]{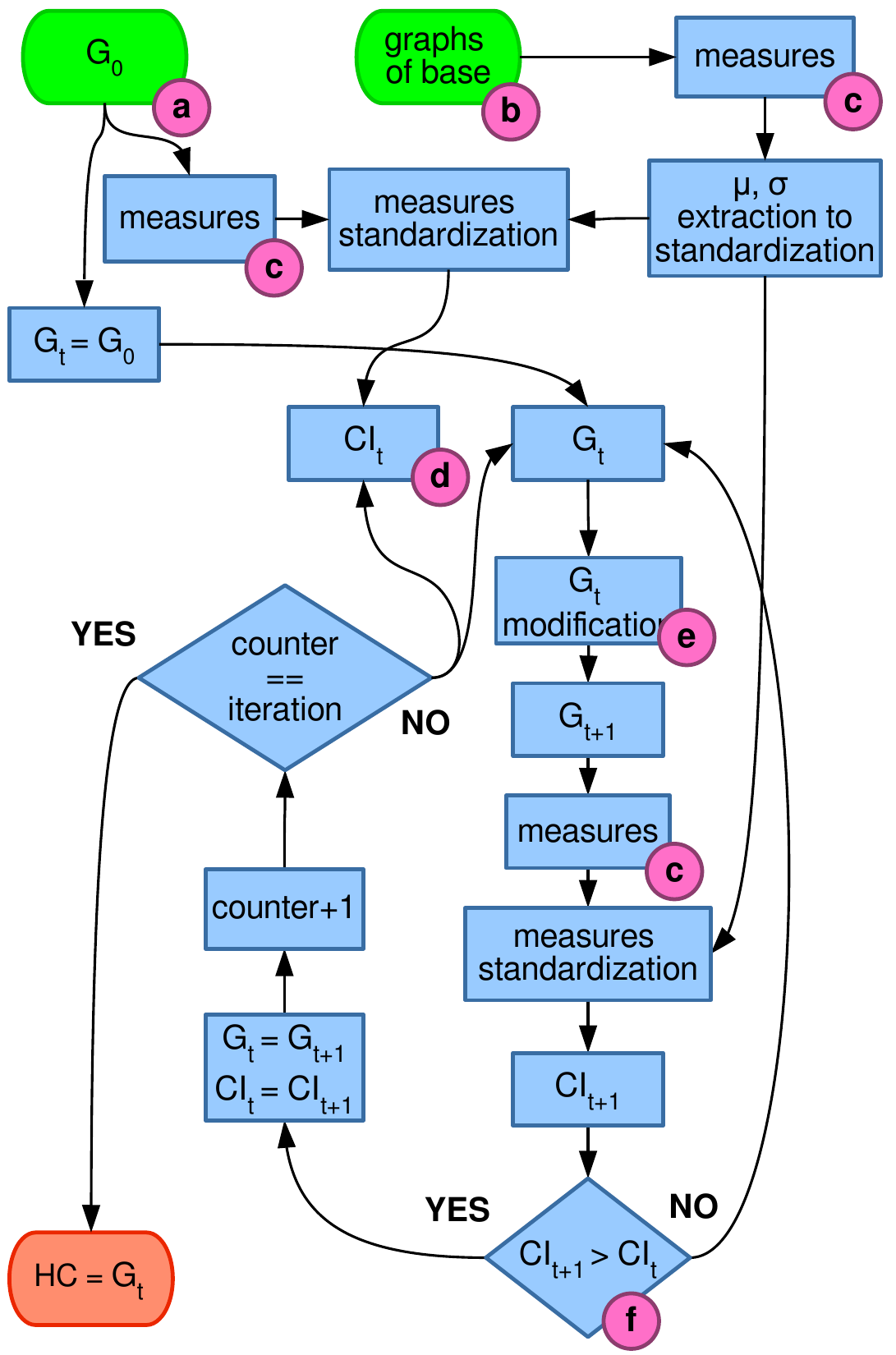}
\caption{\textbf{Hypercomplex Networks Generator.} We approach the algorithm to generate hypercomplex networks as a complexity maximization through network topological changes. In order to do that, we need to define six parameters: \textbf{(a)} Choose the network to have its complexity maximized, $G_0$; \textbf{(b)} Select a set of networks to be used as a reference for comparison with the HC; \textbf{(c)} Pick the measurements to be used to calculate the complexity of the networks; \textbf{(d)} Select the complexity measurement to be maximized. In this work we define \textit{CI} for this purpose, but other quantities can be used in its place; \textbf{(e)} Choose the topological modification to be applied to the network aiming at increasing its complexity; \textbf{(f)} Select the optimization algorithm to maximize the complexity. It is important to notice that the selection of one parameter can influence on the selection of others, for instance the \textit{CI} adopted here assumes the measurements to be standardized.}
\label{fig:hc_generation.pdf}
\end{figure}

It is interesting to approach our search for highly complex networks as an optimization taking place over a given domain, in the specific case of this work a considerably large space of possible reconfigurations of the network edges. As we seek for networks with maximum complexity, the complexity index $CI(G)$ of a network $G$, calculated at each optimization step, is adopted as the function to be maximized. The configuration space is explored through random incremental changes in the network topology. The approach to changing the connections consists of simply choosing an edge $(i,j)$ randomly (with uniform probability), deleting the $j$ node, and then connecting the node $i$ with another node also chosen with uniform probability while not allowing self-loops, connections with nodes that are already neighbors, as well as not leaving $j$ disconnected. 

Gradient descent can be used to implement the sought optimization. The complexity of the new graph, $G_{t+1}$, is compared with the current one, $G_t$, as can be seen in Figure \ref{fig:hc_generation.pdf}. If the complexity of the new configuration is larger, it becomes the current graph and the process is repeated. Otherwise, the candidate configuration is ignored and the processing is repeated. Two stop conditions are set. The principal one is the number of new graphs that are accepted. The second is aimed at limiting the total number of iterations in order to avoid that the algorithm runs indefinitely. The resulting networks are called \textit{hypercomplex networks} -- HCs. The algorithm, as well as the types of choices that can be made, are shown in Figure \ref{fig:hc_generation.pdf}

\section{Results}\label{sec:res}

In order to compare the obtained HC networks with other reference theoretical network models, we considered the ER, BA, Waxman (WAX), Random Geometric Graph (RGG) and Watts-Strogatz (WS) structures having the same number $N=500$ of nodes and same average degree $\langle k \rangle=8$ (with $\pm 5\%$ of tolerance) as the obtained HC networks. In the case of the WS model, it was taken with respect to 2 different reconnecting probabilities, namely $10\%$ and $100\%$. For each one of these models, a total of a hundred graphs is taken into account for the sake of statistical significance. Only the network realizations not including disconnected components have been taken into account.

\begin{figure*}[t!]
\centering

\includegraphics[width=17.5cm]{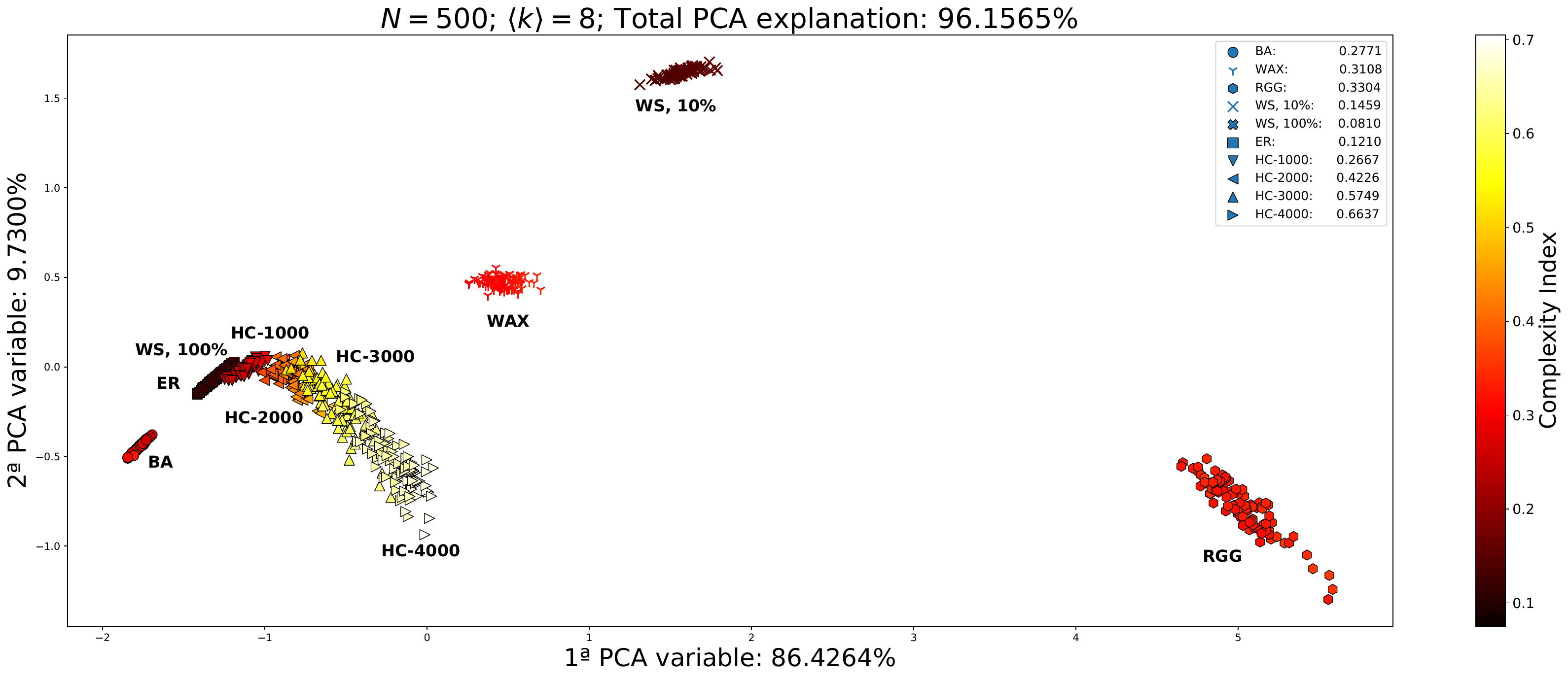}
\caption{The distribution of the several considered networks, including the obtained HC structures as well as traditional theoretical models taken as reference for comparison. The PCA space derives from all 10 respective topological measurements. In the case of the HC model, the networks are mapped with respect to 1000, 2000, 3000 and 4000 taken optimization steps. The heatmap colors have been assigned according to the \textit{CI} value of each network. Several interesting results can be observed, especially the significant separation between the clusters respective to each model and the progressive unfolding of complexity of the HC networks as the number of modifications increases. Another interesting result concerns the fact that the HC networks departs from all other models, including BA, heading to a low density region of the topological features. A substantial increase of the \textit{CI} is observed along the progressively obtained HC networks.}
\label{fig:pca_complexity}
\end{figure*}

The starting points for the HC derivation method were taken as corresponding to ER networks, so that no initial topological biases be implied by these structures thanks to the statistical uniformity of the ER model. For referencing purposes, we have taken the HC networks after 1000, 2000, 3000 and 4000 taken optimization steps, and these networks are henceforth referred to as HC-1000, HC-2000, HC-3000, and HC-4000, respectively.

Principal component analysis (PCA~\cite{wold1987principal, gewers2018principal}) has been applied on all 10 topological measurements of the considered networks so as to project the networks into two dimensions, allowing respective comparative visualization. The results are shown in Figure~\ref{fig:pca_complexity}. The obtained $96.1565\%$ variance explanation corroborates the relevance of the PCA projection. The \textit{CI} respective to each network is also shown in terms of the heatmap.

Several interesting results can be inferred from Figure~\ref{fig:pca_complexity}. First, we have that each of the considered network model yielded a respective cluster that, despite its intrinsic dispersion, is well-separated from the others. Then, and more important, as a consequence of the subsequent optimization steps, the group respective to the HC networks defined an unfolding, starting at the ER models, that is characterized by increasing values of \textit{CI}, toward a previously empty (low density) region in the PCA. Interestingly, the `trajectory' respectively defined tends to move away from all other models, including BA, therefore suggesting that the topology of the HC is distinct from those models. 

\begin{figure}[t]
\centering
\includegraphics[width=8.6cm]{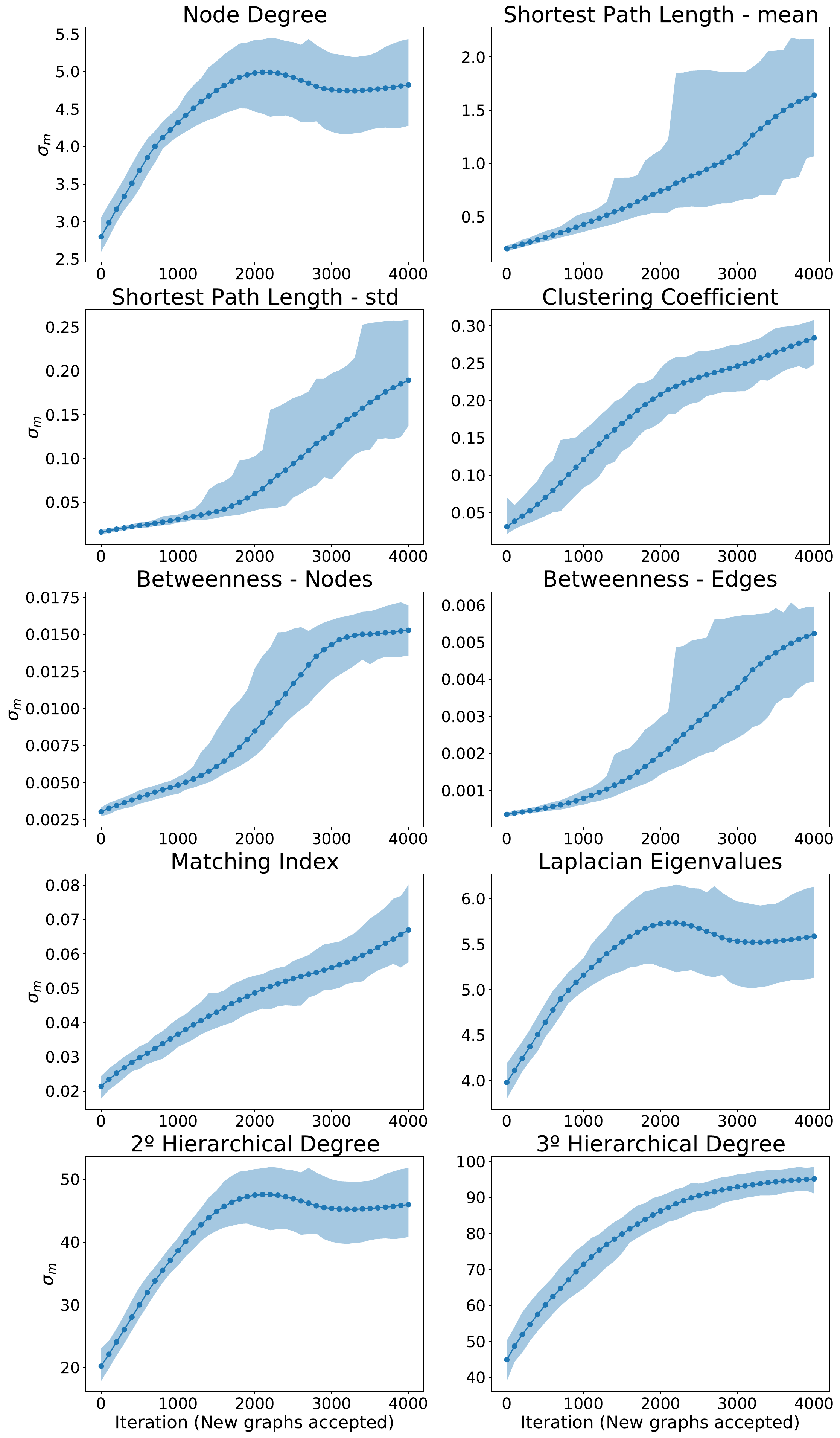}
\caption{Average $\pm$ standard deviation of the several considered topological
measurements of HC networks with $N= 500$ nodes and $\left< k \right> = 8$ along the taken optimization steps (interaction). Observe also that the dispersion of the node degree (as well as the Laplacian eigenvalues and second hierarchical degree) tended to saturate and even decrease, indicating that the complexity of the obtained network could not be obtained by considering only this measurement.}
\label{fig:measures_evolution_hc_4000}
\end{figure}

The obtained results corroborate not only that the HC model is more complex than the other considered types of networks (according to the adopted \textit{CI}), but also that the HC structures displace themselves towards a configuration that is substantially different from the other models. In order to better understand the subsequent changes in the properties associated to the HC model, we present in Figure~\ref{fig:measures_evolution_hc_4000} the average $\pm$ standard deviation of each considered topological measurement along the optimization steps applied to obtain the HC networks. 

\begin{figure}[!t]
\centering
\includegraphics[width=7cm]{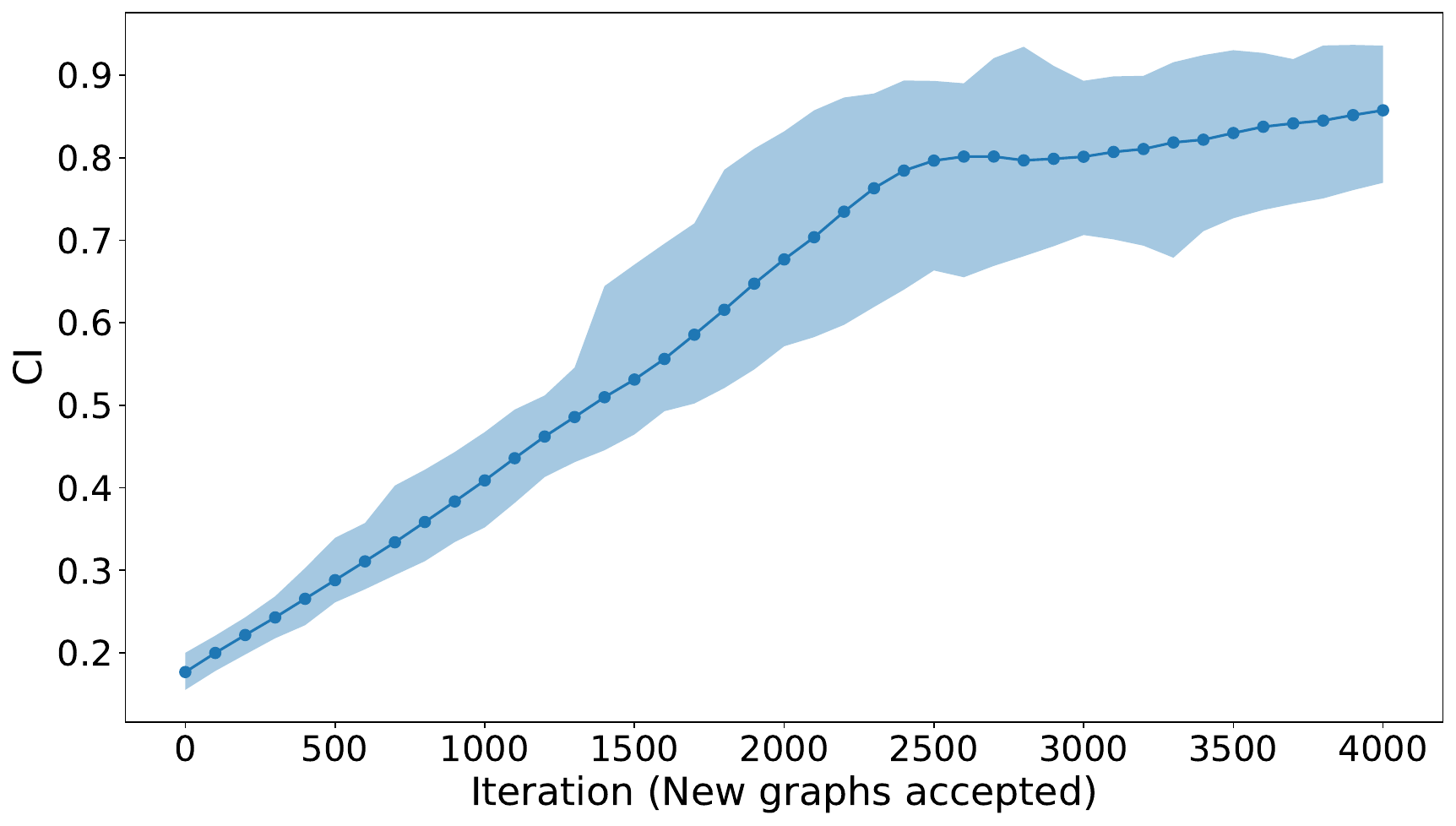}
\caption{Average $\pm$ standard deviation of the \textit{CI} observed during the 
optimization procedure adopted for generating HC networks with $N= 500$ nodes and $\left< k \right> = 8$ in terms of the taken optimization steps (interaction).
Given that the degree distribution obtained for the HC networks is only slightly
broader than that of the ER counterparts, we can conclude that the enhanced complexity of the HC network stems not only from the degree distribution, but also from the heterogeneity of all other considered topological measurements, which have indeed been quantitatively observed to have their respective \textit{CV}s increasing with the interactions.}
\label{fig:complexity_index_steps}
\end{figure}

Interestingly, the optimization dynamics tends to increase the standard deviation respectively to \emph{most} of the adopted measurements, corroborating its effectiveness in deriving networks that are more complex respectively to the overall set of topological features, not only the degree, as can be seen in Figure~\ref{fig:measures_evolution_hc_4000}. The \textit{CI} observed for 4000 realizations of the optimizations is shown in Figure~\ref{fig:complexity_index_steps}, also increasing with the optimization steps.

\begin{figure*}[h!]
\centering
\includegraphics[scale=0.9]{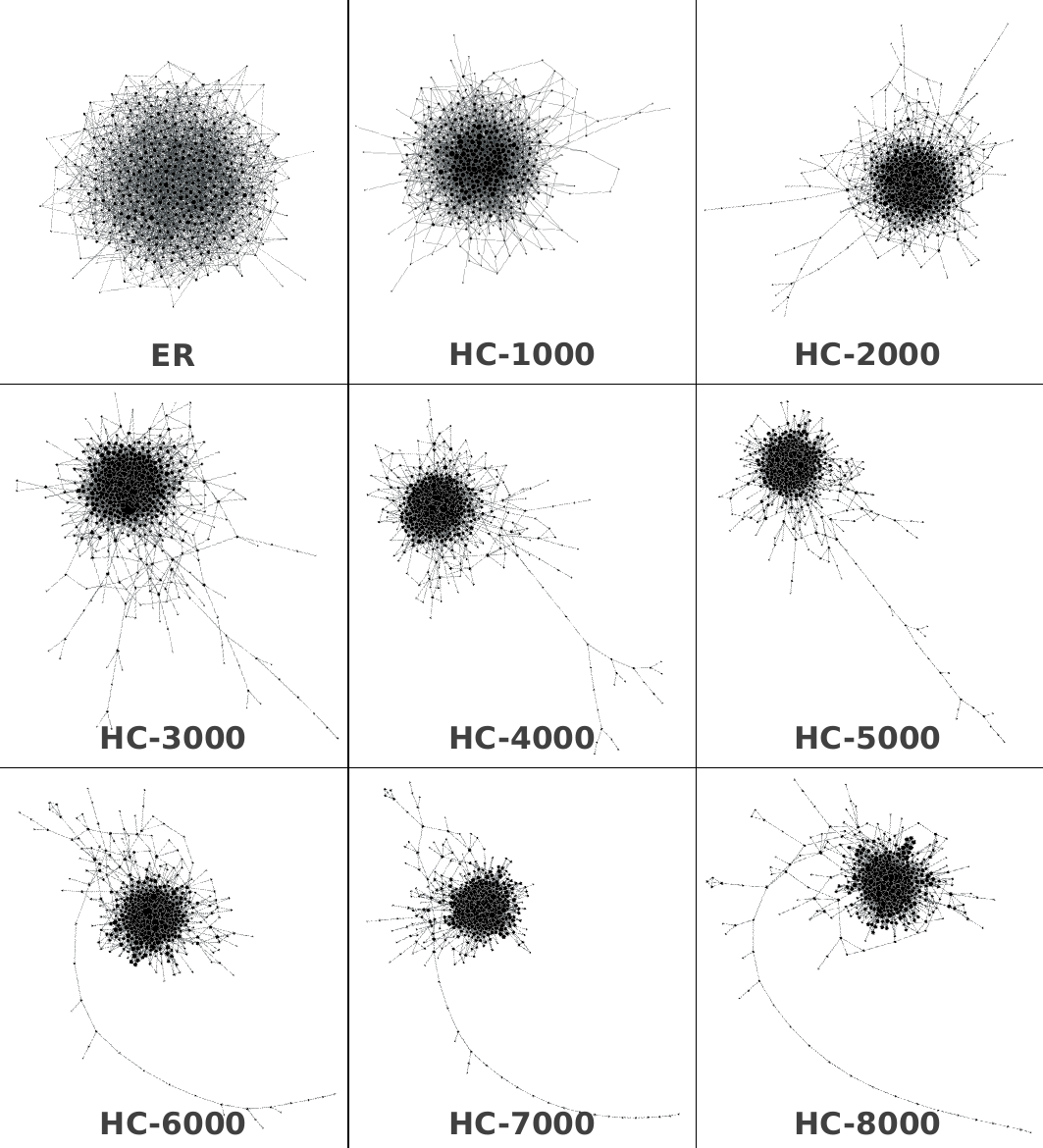}
\caption{An HC network with for $N=500$ nodes and average degree $\left< k \right> = 8$ that was allowed to undergo optimization up to 8000 steps. The diameter of the
nodes has been represented as being proportional to the
respective degree. New peripheral structure, mostly branching subnetworks, are progressively incorporated.}
\label{fig:hc_all_8000}
\end{figure*}

Figure~\ref{fig:hc_all_8000} presents the visualizations of the HC evolution with $N=500$ nodes and $\left< k \right> = 8$, from ER to optimization stage 8000.
Some interesting features can be identified in this figure, including low clustering coefficient, at least at its periphery, as well as a gradient of degree extending from the 
network center toward its periphery. Also interesting is the
tree-like structure appearing at the border of the network.
These obtained structural complexities were absent from the
original ER structures.

\section{Conclusions}\label{sec:concl}

To a good extent, the area of network science derives its motivation from complexity studies, such as the identification of the BA model as a network presenting node degree heterogeneity, more specifically nodes degree distributions following a power law. Indeed, the very concept of \emph{complex} network stems from comparisons with relatively `simple', homogeneous counterparts such as the ER model whose nodes tend to have degree values similar to the average node degree, indicating a kind of statistical node regularity.

In the present work, we aimed at further enhancing the complexity of networks. In order to do so, we started from the fact that the complexity of a network derives not only from heterogeneous degree distributions, but also of present heterogeneity of other, preferably most, other topological measurements~\cite{costa2018complex, kim2008complex}. First, we defined an overall index reflecting the heterogeneity of a large number of topological measurements taken on the given network. More specifically, this complexity index \textit{CI} expresses the relative dispersion of each considered measurement with respect to their maximum value, so that it reaches its maximum value when all measurements have maximum dispersion. Then, we apply an optimization procedure, starting with a uniformly random network (ER), and performing gradient descent over the negative of the respectively obtained \textit{CI}s. The obtained networks, which we have called \emph{hypercomplex networks} -- HC, are thus expected to be substantially more complex than the initial ER networks from which they derived.

Several interesting results have been obtained. More specifically, we have that, as the number of optimization modifications are taken, the HC networks tend to perform a well-defined trajectory in the PCA obtained from several respective topological measurements. This trajectory, which is indeed more similar to elongated clusters, has been observed to head toward a low density position in the PCA space, departing from all the other considered theoretical network models. As expected, the \textit{CI}s of these networks tend to increase progressively. 

The obtained HCs present some additional surprising topological properties such as larger diversity of degree and peripheral branches,  properties nonexistent in the comparative network models. However, the degree distribution obtained for the HCs is only slightly wider than that obtained for ER counterparts. Thus, in a sense the HC networks are not much more complex than ER networks from the perspective of degree distribution. However, at the same time HCs tend to present topology different from all other models, and are characterized by the largest obtained \textit{CI}s. This indicates that the complexity of HCs, as had been initially postulated, derives from the heterogeneity not only of node degree, but also of the other considered topological measurements.

The reported work paves the way to a number of related developments, many of which are currently being pursued. First, it would be interesting to start from theoretical complex network models other than ER. It could be expected that these other models, by being less homogeneous, could imply an initial bias that may or may not correspond to that to be taken by the HCs. In the former case, the changes toward maximum complexity would therefore be delayed so that the initial bias could be first overcome. Another interesting research line concerns the study of the surprising appearance of peripheral branching subnetworks, provided a substantially large number of steps is allowed. These new structures probably imply in an abrupt change of the trajectory underwent by the HCs in the PCA space. It remains a subject of great interest to further study this second modification stage presented by the HCs. 

Another interesting aspect would be to further study the convergence of specific topological features, such as the degree and its hierarchies, as they may indicate critical steps along the optimization dynamics. Given that directionality and weights are known to strongly influence not only the network structure, but also respective dynamics undergoing in these networks, it would be interesting to extend the present work to incorporate those types of networks. 

It would also be interesting to consider alternative complexity indices. Indeed, though the adopted \textit{CI} reflects the overall heterogeneity of the network topological features. Yet another promising possibility is to compare the HC networks with real-world structures, in order not only to try to identify analogue situations, but also to better understand, in a comparative manner, their relative overall complexity.

\section*{Acknowledgments}

Luciano da F. Costa thanks CNPq (grant no.~307085/2018-0) for sponsorship. This work has benefited from FAPESP grant 15/22308-2. \'Everton F. da Cunha thanks CNPq (grant 830717/1999-4 --- 134181/2019-0) for financial support.

\bibliography{refs}

\begin{thebibliography}{16}%
\makeatletter
\providecommand \@ifxundefined [1]{%
 \@ifx{#1\undefined}
}%
\providecommand \@ifnum [1]{%
 \ifnum #1\expandafter \@firstoftwo
 \else \expandafter \@secondoftwo
 \fi
}%
\providecommand \@ifx [1]{%
 \ifx #1\expandafter \@firstoftwo
 \else \expandafter \@secondoftwo
 \fi
}%
\providecommand \natexlab [1]{#1}%
\providecommand \enquote  [1]{``#1''}%
\providecommand \bibnamefont  [1]{#1}%
\providecommand \bibfnamefont [1]{#1}%
\providecommand \citenamefont [1]{#1}%
\providecommand \href@noop [0]{\@secondoftwo}%
\providecommand \href [0]{\begingroup \@sanitize@url \@href}%
\providecommand \@href[1]{\@@startlink{#1}\@@href}%
\providecommand \@@href[1]{\endgroup#1\@@endlink}%
\providecommand \@sanitize@url [0]{\catcode `\\12\catcode `\$12\catcode
  `\&12\catcode `\#12\catcode `\^12\catcode `\_12\catcode `\%12\relax}%
\providecommand \@@startlink[1]{}%
\providecommand \@@endlink[0]{}%
\providecommand \url  [0]{\begingroup\@sanitize@url \@url }%
\providecommand \@url [1]{\endgroup\@href {#1}{\urlprefix }}%
\providecommand \urlprefix  [0]{URL }%
\providecommand \Eprint [0]{\href }%
\providecommand \doibase [0]{http://dx.doi.org/}%
\providecommand \selectlanguage [0]{\@gobble}%
\providecommand \bibinfo  [0]{\@secondoftwo}%
\providecommand \bibfield  [0]{\@secondoftwo}%
\providecommand \translation [1]{[#1]}%
\providecommand \BibitemOpen [0]{}%
\providecommand \bibitemStop [0]{}%
\providecommand \bibitemNoStop [0]{.\EOS\space}%
\providecommand \EOS [0]{\spacefactor3000\relax}%
\providecommand \BibitemShut  [1]{\csname bibitem#1\endcsname}%
\let\auto@bib@innerbib\@empty
\bibitem [{\citenamefont {Costa}\ \emph {et~al.}(2007)\citenamefont {Costa},
  \citenamefont {Rodrigues}, \citenamefont {Travieso},\ and\ \citenamefont
  {Villas~Boas}}]{costa2007characterization}%
  \BibitemOpen
  \bibfield  {author} {\bibinfo {author} {\bibfnamefont {L.~d.~F.}\
  \bibnamefont {Costa}}, \bibinfo {author} {\bibfnamefont {F.~A.}\ \bibnamefont
  {Rodrigues}}, \bibinfo {author} {\bibfnamefont {G.}~\bibnamefont {Travieso}},
  \ and\ \bibinfo {author} {\bibfnamefont {P.~R.}\ \bibnamefont
  {Villas~Boas}},\ }\href@noop {} {\bibfield  {journal} {\bibinfo  {journal}
  {Advances in physics}\ }\textbf {\bibinfo {volume} {56}},\ \bibinfo {pages}
  {167} (\bibinfo {year} {2007})}\BibitemShut {NoStop}%
\bibitem [{\citenamefont {Erd{\H{o}}s}\ and\ \citenamefont
  {R{\'e}nyi}(1959)}]{erdos1959random}%
  \BibitemOpen
  \bibfield  {author} {\bibinfo {author} {\bibfnamefont {P.}~\bibnamefont
  {Erd{\H{o}}s}}\ and\ \bibinfo {author} {\bibfnamefont {A.}~\bibnamefont
  {R{\'e}nyi}},\ }\href@noop {} {\bibfield  {journal} {\bibinfo  {journal}
  {Publicationes Mathematicae}\ }\textbf {\bibinfo {volume} {6}},\ \bibinfo
  {pages} {18} (\bibinfo {year} {1959})}\BibitemShut {NoStop}%
\bibitem [{\citenamefont {Erd{\H{o}}s}\ and\ \citenamefont
  {R{\'e}nyi}(1960)}]{erdos1960evolution}%
  \BibitemOpen
  \bibfield  {author} {\bibinfo {author} {\bibfnamefont {P.}~\bibnamefont
  {Erd{\H{o}}s}}\ and\ \bibinfo {author} {\bibfnamefont {A.}~\bibnamefont
  {R{\'e}nyi}},\ }\href@noop {} {\bibfield  {journal} {\bibinfo  {journal}
  {Publ. Math. Inst. Hung. Acad. Sci}\ }\textbf {\bibinfo {volume} {5}},\
  \bibinfo {pages} {17} (\bibinfo {year} {1960})}\BibitemShut {NoStop}%
\bibitem [{\citenamefont {Barab{\'a}si}\ and\ \citenamefont
  {Albert}(1999)}]{barabasi1999emergence}%
  \BibitemOpen
  \bibfield  {author} {\bibinfo {author} {\bibfnamefont {A.-L.}\ \bibnamefont
  {Barab{\'a}si}}\ and\ \bibinfo {author} {\bibfnamefont {R.}~\bibnamefont
  {Albert}},\ }\href@noop {} {\bibfield  {journal} {\bibinfo  {journal}
  {science}\ }\textbf {\bibinfo {volume} {286}},\ \bibinfo {pages} {509}
  (\bibinfo {year} {1999})}\BibitemShut {NoStop}%
\bibitem [{\citenamefont {Kim}\ and\ \citenamefont
  {Wilhelm}(2008)}]{kim2008complex}%
  \BibitemOpen
  \bibfield  {author} {\bibinfo {author} {\bibfnamefont {J.}~\bibnamefont
  {Kim}}\ and\ \bibinfo {author} {\bibfnamefont {T.}~\bibnamefont {Wilhelm}},\
  }\href@noop {} {\bibfield  {journal} {\bibinfo  {journal} {Physica A:
  Statistical Mechanics and its Applications}\ }\textbf {\bibinfo {volume}
  {387}},\ \bibinfo {pages} {2637} (\bibinfo {year} {2008})}\BibitemShut
  {NoStop}%
\bibitem [{\citenamefont {Costa}(2018)}]{costa2018complex}%
  \BibitemOpen
  \bibfield  {author} {\bibinfo {author} {\bibfnamefont {L.~d.~F.}\
  \bibnamefont {Costa}},\ }\href@noop {} {\  (\bibinfo {year} {2018})},\
  \bibinfo {note}
  {\url{https://www.researchgate.net/publication/324312765_What_is_a_Complex_Network_CDT-2}.
  [Online; accessed 07-October-2020.]}\BibitemShut {NoStop}%
\bibitem [{\citenamefont {Watts}\ and\ \citenamefont
  {Strogatz}(1998)}]{watts1998collective}%
  \BibitemOpen
  \bibfield  {author} {\bibinfo {author} {\bibfnamefont {D.~J.}\ \bibnamefont
  {Watts}}\ and\ \bibinfo {author} {\bibfnamefont {S.~H.}\ \bibnamefont
  {Strogatz}},\ }\href@noop {} {\bibfield  {journal} {\bibinfo  {journal}
  {Nature}\ }\textbf {\bibinfo {volume} {393}},\ \bibinfo {pages} {440}
  (\bibinfo {year} {1998})}\BibitemShut {NoStop}%
\bibitem [{\citenamefont {Freeman}(1977)}]{freeman1977set}%
  \BibitemOpen
  \bibfield  {author} {\bibinfo {author} {\bibfnamefont {L.~C.}\ \bibnamefont
  {Freeman}},\ }\href@noop {} {\bibfield  {journal} {\bibinfo  {journal}
  {Sociometry}\ ,\ \bibinfo {pages} {35}} (\bibinfo {year} {1977})}\BibitemShut
  {NoStop}%
\bibitem [{\citenamefont {Hilgetag}\ \emph {et~al.}(2002)\citenamefont
  {Hilgetag}, \citenamefont {K{\"o}tter}, \citenamefont {Stephan},\ and\
  \citenamefont {Sporns}}]{hilgetag2002computational}%
  \BibitemOpen
  \bibfield  {author} {\bibinfo {author} {\bibfnamefont {C.~C.}\ \bibnamefont
  {Hilgetag}}, \bibinfo {author} {\bibfnamefont {R.}~\bibnamefont
  {K{\"o}tter}}, \bibinfo {author} {\bibfnamefont {K.~E.}\ \bibnamefont
  {Stephan}}, \ and\ \bibinfo {author} {\bibfnamefont {O.}~\bibnamefont
  {Sporns}},\ }in\ \href@noop {} {\emph {\bibinfo {booktitle} {Computational
  Neuroanatomy}}}\ (\bibinfo  {publisher} {Springer},\ \bibinfo {year} {2002})\
  pp.\ \bibinfo {pages} {295--335}\BibitemShut {NoStop}%
\bibitem [{\citenamefont {de~Arruda}\ \emph {et~al.}(2018)\citenamefont
  {de~Arruda}, \citenamefont {Silva}, \citenamefont {Marinho}, \citenamefont
  {Amancio},\ and\ \citenamefont {Costa}}]{arruda2018representation}%
  \BibitemOpen
  \bibfield  {author} {\bibinfo {author} {\bibfnamefont {H.~F.}\ \bibnamefont
  {de~Arruda}}, \bibinfo {author} {\bibfnamefont {F.~N.}\ \bibnamefont
  {Silva}}, \bibinfo {author} {\bibfnamefont {V.~Q.}\ \bibnamefont {Marinho}},
  \bibinfo {author} {\bibfnamefont {D.~R.}\ \bibnamefont {Amancio}}, \ and\
  \bibinfo {author} {\bibfnamefont {L.~d.~F.}\ \bibnamefont {Costa}},\
  }\href@noop {} {\bibfield  {journal} {\bibinfo  {journal} {Journal of Complex
  Networks}\ }\textbf {\bibinfo {volume} {6}},\ \bibinfo {pages} {125}
  (\bibinfo {year} {2018})}\BibitemShut {NoStop}%
\bibitem [{\citenamefont {Seary}\ and\ \citenamefont
  {Richards}(1995)}]{seary1995partitioning}%
  \BibitemOpen
  \bibfield  {author} {\bibinfo {author} {\bibfnamefont {A.~J.}\ \bibnamefont
  {Seary}}\ and\ \bibinfo {author} {\bibfnamefont {W.~D.}\ \bibnamefont
  {Richards}},\ }in\ \href@noop {} {\emph {\bibinfo {booktitle} {Proceedings of
  the International Conference on Social Networks}}},\ Vol.~\bibinfo {volume}
  {1}\ (\bibinfo {year} {1995})\ pp.\ \bibinfo {pages} {47--58}\BibitemShut
  {NoStop}%
\bibitem [{\citenamefont {Newman}(2006)}]{newman2006finding}%
  \BibitemOpen
  \bibfield  {author} {\bibinfo {author} {\bibfnamefont {M.~E.}\ \bibnamefont
  {Newman}},\ }\href@noop {} {\bibfield  {journal} {\bibinfo  {journal}
  {Physical review E}\ }\textbf {\bibinfo {volume} {74}},\ \bibinfo {pages}
  {036104} (\bibinfo {year} {2006})}\BibitemShut {NoStop}%
\bibitem [{\citenamefont {Newman}(2018)}]{newman2018networks}%
  \BibitemOpen
  \bibfield  {author} {\bibinfo {author} {\bibfnamefont {M.}~\bibnamefont
  {Newman}},\ }\href@noop {} {\emph {\bibinfo {title} {Networks}}}\ (\bibinfo
  {publisher} {Oxford university press},\ \bibinfo {year} {2018})\BibitemShut
  {NoStop}%
\bibitem [{\citenamefont {Costa}\ and\ \citenamefont
  {Silva}(2006)}]{costa2006hierarchical}%
  \BibitemOpen
  \bibfield  {author} {\bibinfo {author} {\bibfnamefont {L.~d.~F.}\
  \bibnamefont {Costa}}\ and\ \bibinfo {author} {\bibfnamefont {F.~N.}\
  \bibnamefont {Silva}},\ }\href@noop {} {\bibfield  {journal} {\bibinfo
  {journal} {Journal of Statistical Physics}\ }\textbf {\bibinfo {volume}
  {125}},\ \bibinfo {pages} {841} (\bibinfo {year} {2006})}\BibitemShut
  {NoStop}%
\bibitem [{\citenamefont {Wold}\ \emph {et~al.}(1987)\citenamefont {Wold},
  \citenamefont {Esbensen},\ and\ \citenamefont {Geladi}}]{wold1987principal}%
  \BibitemOpen
  \bibfield  {author} {\bibinfo {author} {\bibfnamefont {S.}~\bibnamefont
  {Wold}}, \bibinfo {author} {\bibfnamefont {K.}~\bibnamefont {Esbensen}}, \
  and\ \bibinfo {author} {\bibfnamefont {P.}~\bibnamefont {Geladi}},\
  }\href@noop {} {\bibfield  {journal} {\bibinfo  {journal} {Chemometrics and
  intelligent laboratory systems}\ }\textbf {\bibinfo {volume} {2}},\ \bibinfo
  {pages} {37} (\bibinfo {year} {1987})}\BibitemShut {NoStop}%
\bibitem [{\citenamefont {Gewers}\ \emph {et~al.}(2018)\citenamefont {Gewers},
  \citenamefont {Ferreira}, \citenamefont {de~Arruda}, \citenamefont {Silva},
  \citenamefont {Comin}, \citenamefont {Amancio},\ and\ \citenamefont
  {Costa}}]{gewers2018principal}%
  \BibitemOpen
  \bibfield  {author} {\bibinfo {author} {\bibfnamefont {F.~L.}\ \bibnamefont
  {Gewers}}, \bibinfo {author} {\bibfnamefont {G.~R.}\ \bibnamefont
  {Ferreira}}, \bibinfo {author} {\bibfnamefont {H.~F.}\ \bibnamefont
  {de~Arruda}}, \bibinfo {author} {\bibfnamefont {F.~N.}\ \bibnamefont
  {Silva}}, \bibinfo {author} {\bibfnamefont {C.~H.}\ \bibnamefont {Comin}},
  \bibinfo {author} {\bibfnamefont {D.~R.}\ \bibnamefont {Amancio}}, \ and\
  \bibinfo {author} {\bibfnamefont {L.~d.~F.}\ \bibnamefont {Costa}},\
  }\href@noop {} {\bibfield  {journal} {\bibinfo  {journal} {arXiv preprint
  arXiv:1804.02502}\ } (\bibinfo {year} {2018})}\BibitemShut {NoStop}%
\end{thebibliography}%

\end{document}